# Through-membrane electron-beam lithography for ultrathin membrane applications

M. Neklyudova, A. K. Erdamar, L. Vicarelli, S. J. Heerema, T. Rehfeldt, G. Pandraud, Z. Kolahdouz, C. Dekker, and H. W. Zandbergen

# Through-membrane electron-beam lithography for ultrathin membrane applications

M. Neklyudova,[1] A. K. Erdamar,[1] L. Vicarelli,[1] S. J. Heerema,[1] T. Rehfeldt,[1] G. Pandraud,[2] Z. Kolahdouz,[1] C. Dekker,[1] and H. W. Zandbergen[1,a)]

[1]Kavli Institute of Nanoscience, Delft University of Technology, Delft 2628CJ, the Netherlands
[2]Else Kooi Laboratory, Delft University of Technology, Delft 2628CT, the Netherlands

We present a technique to fabricate ultrathin (down to 20 nm) uniform electron transparent windows at dedicated locations in a SiN membrane for *in situ* transmission electron microscopy experiments. An electron-beam (e-beam) resist is spray-coated on the backside of the membrane in a KOH-etched cavity in silicon which is patterned using through-membrane electron-beam lithography. This is a controlled way to make transparent windows in membranes, whilst the topside of the membrane remains undamaged and retains its flatness. Our approach was optimized for MEMS-based heating chips but can be applied to any chip design. We show two different applications of this technique for (1) fabrication of a nanogap electrode by means of electromigration in thin free-standing metal films and (2) making low-noise graphene nanopore devices. *Published by AIP Publishing.*
[http://dx.doi.org/10.1063/1.4986991]

For *in situ* TEM heating and biasing experiments, one needs chips containing less than 20 nm-thick amorphous membrane windows to obtain a good electron transparency.[1,2] Once the freestanding area of the membrane needs to be larger than several microns, or when the membrane design is more complex (i.e., includes metal electrodes), it becomes very difficult to maintain the membranes intact. One way to obtain these thin membrane windows is to pattern a mask (such as a resist layer) on the topside of the SiN and etch the material down to the Si, followed by deposition of a thin layer of SiN using low pressure chemical vapor deposition (LPCVD). This approach has one big disadvantage: the thin windows are formed at the bottom of the thick SiN membrane (see Fig. S1 in the supplementary material), meaning that the SiN window is in a deep cavity compared to the top surface of the membrane. Such deep cavity-shaped membrane windows are very inconvenient for various measurements, such as liquid *in situ* TEM experiments. Also, additional deposition of electrical contacts for biasing *in situ* TEM experiments on these recessed surfaces is impossible. To address such issues, we have developed a technique where the SiN is locally removed from the backside, such that the topside of the membrane remains undamaged.

Performing electron-beam lithography (EBL) from the backside of the membrane is not trivial with conventional lithography systems since it stands at the bottom of a 300–500 $\mu$m-deep KOH-etched cavity. The resist mask should cover the surface that is going to be patterned and with the distance of less than 100 $\mu$m to the lens of the EBL device to avoid defocusing issues.

In this paper, we show that the e-beam resist, which is spray coated on the backside of the SiN membrane, can be exposed with an e-beam from the topside and, therefore, through the membrane and the structures on top of it. In this way, the defocusing problem is solved, since the lens will be in close enough proximity to the resist layer. After explanation of the full procedure, we present two examples of applications of this approach.

Fabrication of the MEMS-based heaters consists of several steps. First, we start with 300–500 $\mu$m thick Si wafers and deposit 200 nm-thick SiN by the low pressure chemical vapor deposition (LPCVD) technique as an isolation layer between the metal and the Si substrate. Heater coils are made of Tantalum (Ta)/Platinum (Pt) metal layers with 20/180 nm thicknesses, respectively, deposited by e-beam evaporation and etched by ion etching. After the second deposition of 200 nm-thick LPCVD SiN, the heater is embedded in a total 400 nm-thick SiN. The free-standing SiN membranes are obtained by KOH-etching of Si.

Next, e-beam resist (PMMA:PGMEA:MEK) is sprayed in multiple steps on the backside of the chips with an EVG101 spray-coater. The main challenge of this method is to obtain uniform resist coverage on the backside of SiN membranes due to high topography of the 300–500 $\mu$m-deep KOH-etched cavity (shown schematically in Fig. 1). The optimized recipe results in a 12 $\mu$m-thick PMMA resist layer. The resist is exposed from the topside of the chip by 100 kV electrons penetrating through the 400 nm-thick SiN membrane

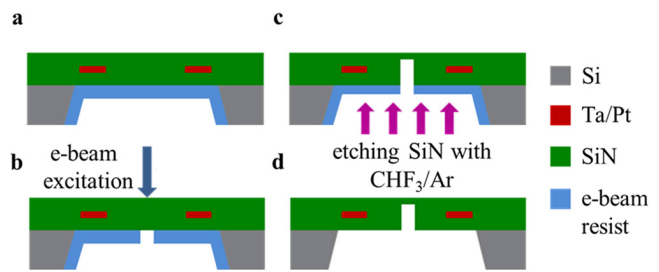

FIG. 1. Schematic illustration (not to scale) of the fabrication process of backside opening by excitation from the top. (a) Spray-coating of e-beam resist from the backside of the chip, (b) EBL exposure and development, (c) RIE of SiN from the back, up to required thickness, and (d) after removal of the resist by PRS3000 (positive resist stripper).

a)Electronic mail: H.W.Zandbergen@tudelft.nl

with a Leica 5000+ EBL machine. After resist development, the membrane is etched by reactive ion etching (RIE) in CHF$_3$/Ar-based plasma with anisotropic etching. The etching rate and time are critical to obtain the required thickness of the SiN in RIE. Here, we used a Leybold RIE machine at 50 W (Ar:CHF$_3$ 25:25 sccm) to obtain an etch rate of 18 nm/min.

The membrane thickness is monitored using optical microscopy, where the SiN thickness is estimated based on comparison of SiN color with LPCVD SiN color chart. After etching, the resist is removed by PRS3000 (positive resist stripper) and O$_2$-plasma.

Depending on the application of this technique, additional fabrication steps might be required.

The first example of an application of through-membrane lithography technique is for *in situ* TEM biasing experiments, in particular, the fabrication of a nanogap electrode in gold (Au) nanobridges by electromigration.

Nanogap electrodes represent a pair of electrodes separated with a nanogap of only a few nanometers (1–10 nm). Nanogap electrodes are essential tools for characterization of material properties at the nanometer scale and used for fabrication of molecular-scale devices and circuits.[3] Nanogap electrodes can be fabricated by different methods such as mechanical break junctions,[4] EBL,[5] feedback controlled electromigration (FCE),[6] and shadow mask evaporation.[7] All these methods show promising results and provide a desired configuration of the electrodes. In this work, we show the fabrication of nanogap electrodes in an Au nanobridge by the FCE technique while monitoring this *in situ* with TEM. *In situ* TEM allows the direct observation of nanogap formation in real-time and possible control of the final size of the nanogap.[8]

The through-membrane EBL is used to fabricate free-standing metallic nanobridges on top of the heating chips used for *in situ* TEM heating experiments (see Fig. 2). The free-standing configuration of the nanobridges with a subsequent shaping into a nanogap electrode is required for the characterization of subsequently trapped nanospecies, because a thick SiN support leads to a too noisy background for TEM visualization of specimen behavior under an applied stimulus. The interest to fabricate these nanogap electrodes on top of the heaters is based on the feasibility of performing TEM characterization of nanomaterials (low-dimensional nanocrystals, phase change nanoparticles, individual molecules, etc.), while applying both voltage and heating simultaneously. In our case, heating up to 120–140 °C allows preventing the e-beam-induced carbon contamination during *in situ* visualization of nanogap formation.

A schematic illustration of MEMS-based heater with a flat 400 nm-thick SiN center is shown in Fig. 2(a). Through-membrane EBL with following RIE was applied to obtain 50 nm-thick SiN windows with 5 μm in diameter in the thick SiN membrane in the center of Pt heating coil [see Fig. 2(b)]. Au bridges with a length of 700 nm, a width of 250 nm, and a thickness of 20 nm were made on top of 50 nm-thick SiN area using EBL followed by e-beam evaporation from the Au source. Contact pads to the bridge were placed by the second step of EBL and metal evaporation of a 250 nm-thick layer of Au on a 5 nm-thick adhesion layer of Cr. The configuration of the fabricated device is schematically presented in Fig. 2(c). To further reduce the thickness of SiN at the location of the Au nanobridge, SiN was etched from the backside of MEMS-based heater using RIE with CHF$_3$/O$_2$ gases with a flow ratio of 50 sccm and 2.5 sccm, respectively.

The MEMS-based heater with Au nanobridge was placed into a built-in-house TEM holder containing six contacts, which allows the combination of heating experiments and electrical measurements. Four contacts are used for heating. Electrical measurements are done by applying voltage to the remaining two contacts. The heater spiral was calibrated with a pyrometer before the experiment.

The FCE process in Au nanobridges was investigated by *in situ* TEM using a FEI Titan microscope operating at 300 keV. In the FCE mode, the bridge conductance was constantly monitored while the voltage was ramping up. If there was a sudden decrease in conductance, the voltage was reduced to a lower value. The process started again after a new reference conductance was defined. The program stopped when a pre-defined conductance value was reached. To avoid e-beam-induced carbon contamination, which can result in spurious conductance, the electromigration experiments were performed at a temperature of 120 °C.

Figure 3 shows snapshots of a typical *in situ* TEM movie recorded during the nanogap formation in the FCE process. The initial configuration of the bridge is shown in Fig. 3(a). When the current was passed through the bridge, we observed grain growth prior to electromigration. The grain growth occurred due to the temperature rise in the bridge caused by current-induced Joule-heating. In Fig. 3(b), the bridge started to thin close to the cathode side forming a constriction. When the electromigration was observed, the current density was about $7 \times 10^7$ A/cm$^2$. In the last stage of the electromigration, a narrow constriction was formed as shown in Fig. 3(c) and the constriction was further narrowed at low voltage values (around 200 mV). Finally, a nanogap electrode was produced [see Fig. 3(d)]. The size of the just formed nanogap, which was about 2–3 nm, increased to 5–6 nm during continuous illumination by the e-beam within several seconds. Figure 3(e) shows a TEM image acquired at higher magnification of the final configuration of the nanogap with the size of 5.26 nm, which did not change for at least several minutes of e-beam irradiation. At the moment when the nanogap electrode was initially formed, the tips of the electrodes were sharp. However, after several seconds when the gap size increased, the shape of the electrode tips became more smooth, which is likely due to surface tension of Au. This observation is in agreement with the previous reports of Zandbergen *et al.*[9] on continued relaxation of Au nanogaps formed by e-beam bombardment even after the intense irradiation is completed, also with the report of Strachan *et al.*[10] on the evolution of Au nanogap electrodes. A typical I–V curve of the FCE process in Au nanobridges is shown in Fig. 3(f).

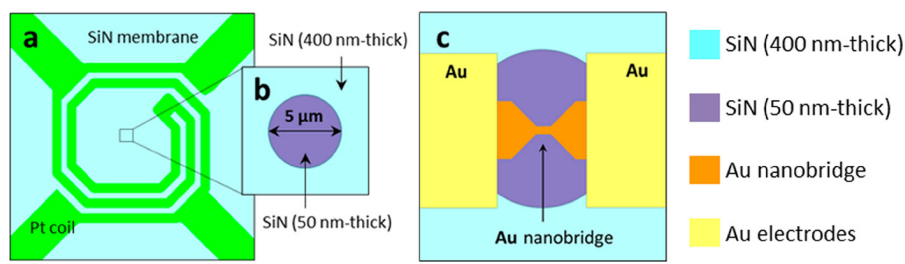

FIG. 2. (a) Schematic illustration of MEMS-based heater with a flat 400 nm-thick SiN center. (b) Central part of the heating coil showing a 50 nm-thick SiN window with a diameter of 5 μm. (c) The configuration of the fabricated device onto the 50 nm-thick SiN window containing the 20 nm-thick Au nanobridge and 200 nm-thick Au contact pads.

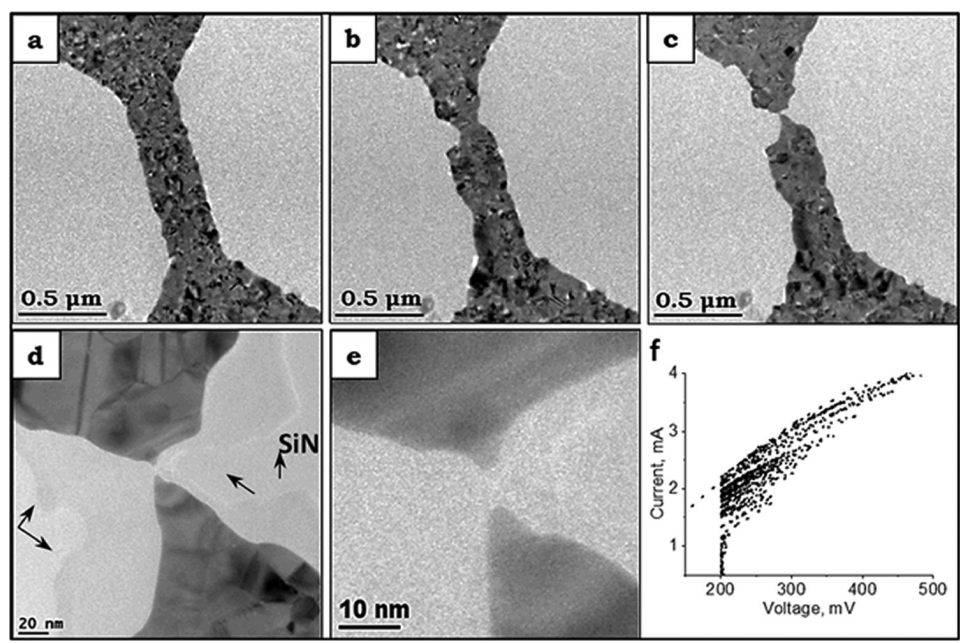

FIG. 3. Snapshots taken from the *in situ* TEM movie recorded during the FCE process in the Au nanobridge. (a) The original bridge. (b) and (c) TEM images showing formation of nanogaps in the Au bridge during FCE. (d) TEM image acquired after nanogap formation; arrows indicate the areas with a thin SiN layer. (e) Enlarged area of (d) showing Au electrodes separated with a nanogap of 5.26 nm. (f) I-V curve of the FCE process.

In Fig. 3(d), one can see a thin layer of SiN remained around the nanogap. Compared to standard fabrication methods of nanogaps on top of SiN membranes with the thicknesses of 100 nm,[11] our technique allows the fabrication of nanogap electrodes with a thin SiN layer beneath it.

To remove the SiN completely, we applied a vaporized HF by SPTS etch vapor system, using a 190 sccm HF flow that enabled to etch remained 2–3 nm SiN completely without damaging the metal lines (see Fig. S2 in the supplementary material). After applying HF treatment, we observed no SiN near the metal structures.

The second example of application of through-membrane lithography technique is for low-noise graphene nanopore devices. Graphene nanopores represent a promising tool for fast and direct sequencing of DNA molecules.[19] In nanopore sensing, a tiny hole ("nanopore") in a membrane that separates two compartments of electrolyte solution is the only pathway for ions and molecules to pass. When a voltage is applied over the membrane, an ionic current is induced with a resistance that is set by the pore length and width, and negatively charged DNA molecules will move towards the positive pole. When a DNA molecule traverses through the nanopore, it impedes the ionic current, which leads to a resistive spike (typically ~1 nA) in the ionic current baseline (several nA). The two great advantages of graphene nanopores, to the more commonly used SiN solid-state nanopores, are that the graphene is atomically thin, which optimizes the sensing resolution as the pore hosts a minimum number of bases at the same time, and its conductive nature facilitates new modes of base detection.

One downside of these graphene nanopore systems is that the noise levels in the ionic current are relatively high (about two orders of magnitude higher than in SiN pores[12]). The noise in the graphene nanopore ionic current is characterized by a 1/f dependence [Fig. 4(c), blue curve]. It has been shown that increasing the number of graphene layers,[13] the use of additional layers of other materials,[13,14] or reducing the area of freestanding graphene can lower the noise levels in the graphene nanopore currents.[15] Reducing the area of freestanding graphene (diameter ~100 nm) provides the most elegant approach to the noise reduction as the atomically thin membrane is retained.

As in the heater chip layout described above, a Pt heater is embedded in the supporting SiN membrane, which is used to heat up the graphene during STEM sculpting of the graphene nanopore,[16,17] leading to a total membrane thickness of 600 nm [Fig. 4(a)]. The formation of a narrow (d~100 nm) access channel in such a "thick" membrane would add a large channel resistance to the circuit. This is an unwanted effect, as in the ideal case the graphene nanopore exclusively sets the resistance and thus behaves as the sensing probe. To reduce the area of freestanding graphene diameter to ~100 nm, while preventing the addition of a huge channel resistance, the fabrication of a thin window in the SiN membrane is needed. Using the through-membrane EBL technique and RIE that is described in this paper, $5 \times 5\ \mu m^2$ windows of ~50–150 nm thickness were fabricated, after which pores with diameters of ~50–150 nm were made using focused ion beam drilling. Graphene flakes were subsequently transferred onto these chips,[18] and finally ~10 nm pores were sculpted in the graphene using high-temperature sculpting with STEM.[16,17]

We measured and analyzed ionic current baselines of 24 thin window devices with reduced areas of freestanding graphene (~50–150 nm in diameter) and compared their noise levels to those of 45 devices with 600 nm-thick windows containing $\sim 1\,\mu m$ diameter freestanding graphene. All traces were recorded at 100 mV at a KCl or LiCl salt concentration of 1 M. Two representative current baselines are plotted in Fig. 4(b). These clearly show that the blue curve originating from a device with a large area of freestanding graphene (d$\sim 1\,\mu m$) is fluctuating more than the red curve belonging to a reduced area of freestanding graphene (d$\sim 100$ nm). This occurs at all bandwidths, but the difference is much more pronounced for a lower bandwidth [cf. the traces in light blue and grey in Fig. 4(b)]. The current power spectral densities that correspond to the traces from Fig. 4(b) are compared in Fig. 4(c). The two types of devices clearly expose different power spectral density curves. The noise from the device with a large area of freestanding graphene (d$\sim 1\,\mu m$) (blue) is characterized by a 1/f dependence up to the filter cut-off frequency of 10 kHz, whereas for the backside etched samples, the 1/f dependence only holds up to a few hundred Hz. To determine the low-frequency noise coefficient ($C_{LF}$) per device, representing the 1/f noise level, the power spectral density functions were normalized by the squares of their mean currents and linear fits to the curves between 1 and 200 Hz were applied on logarithmic scales (see Ref. 12 for more details on the analysis). The results for $C_{LF}$ are plotted in Fig. 4(d); we find that the 1/f noise levels in thin window devices are on average reduced by one order of magnitude [$C_{LF(d\sim 100\,nm)}\sim 5 \times 10^{-7}$ versus $C_{LF(d\sim 1\,\mu m)}\sim 6 \times 10^{-6}$].

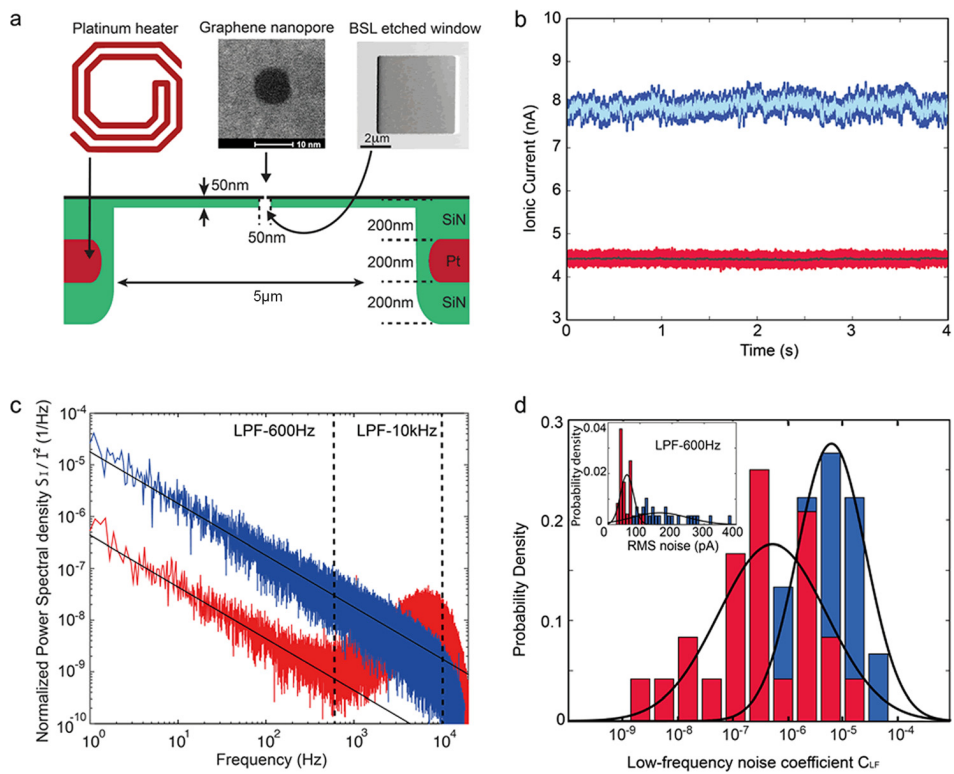

FIG. 4. Graphene nanopores fabricated on thin SiN windows with smaller SiN pore sizes exhibit lower noise due to a reduced area of freestanding graphene. (a) Scheme of a graphene nanopore device with a $5 \times 5\,\mu m^2$ and 50 nm-thick window, etched from the backside of a 600 nm-thick SiN membrane. (b) Ionic current baselines of a device with a large area of freestanding graphene ($\sim 1\,\mu m$ in diameter) on a 600 nm SiN membrane (blue) and of a device with a backside etched window [as in Fig. 3(a)] with a small area of freestanding graphene ($\sim 100$ nm in diameter) (red). (c) Normalized power spectral density curves of the ionic current baselines from Fig. 3(b). The 1/f noise level is determined by linear fitting of the logarithmic values between 1 and 200 Hz (black curves). (d) Probability distributions of the 1/f noise coefficients $C_{LF}$ of the two different device layouts, and the 1/f noise levels in thin window devices are on average reduced by one order of magnitude. Inset: RMS noise levels at a filter frequency of 600 Hz, where the thin window devices with the reduced area of freestanding graphene (red) expose a factor of $\sim 4$ lower RMS noise.

Second, we quantified the $I_{rms}$ levels (representing the deviations from the mean of the current) of 24 thin window devices with reduced freestanding graphene areas (red) and compared those to 29 "thick" membrane samples with a larger freestanding graphene area (blue).

As can be read from the power spectral density curves in Fig. 4(c), the noise is particularly reduced in the low frequency regime ($<600$ Hz), which is observed by the comparison of the light blue and dark grey traces in Fig. 4(b) that were low-pass filtered at 600 Hz. The $I_{rms}$ values at a bandwidth of 600 Hz for both device types are represented in the histogram in the inset in Fig. 4(d), showing that the $I_{rms}$ noise is reduced by a factor of $\sim 4$ (from $146 \pm 16$ pA to $35 \pm 5$ pA). At 10 kHz, the $I_{rms}$ noise is reduced by a factor of $\sim 2.7$ ($162 \pm 16$ pA to $59 \pm 4$).

The noise reduction that we have shown here is highly relevant for DNA sensing with graphene nanopores as it enables improving the signal-to-noise, although other challenges need to be overcome as well. Various research groups are currently exploring alternative detection methods that use the conductive nature of the graphene.[19] In these setups, a good signal-to-noise ratio is also important as the nanopore principle is still used, both to drive the DNA molecule along the sensor and to confirm its passage.

We have shown a fabrication technique so called through-membrane EBL with exposure from the top that enables us to obtain very thin electron transparent SiN windows in a membrane. Exposure of the e-beam resist through 400–600 nm-thick

SiN membranes facilitates the manufacturing of a differnt type of geometry in which one can control and maintain the flatness of the topside of the membrane while creating very thin windows at certain locations. Compared to optical lithography, EBL allows one to obtain smaller structures with a high accuracy down to the nanometer scale. The two applications of through-membrane EBL given in this paper are frontier projects and examples of the potential of this fabrication technique.

In nanogap formation, we have demonstrated that this unique fabrication technique enables us to obtain a very thin SiN layer. The application of additional etching steps (with vaporized HF) allows the removal of the membrane completely for further applications of nanogap electrodes. In the second application, the thin windows are used to fabricate low noise nanopores in graphene to improve the signal-to-noise levels in DNA sensing experiments. This development is beneficial for further graphene nanopore measurements and for graphene-based DNA sequencing.

In addition to these demonstrated applications, having an ultrathin SiN electron transparent window on the top of the membrane will have advantages on *in situ* TEM liquid cell, nanoreactor, and battery studies since it offers a flat surface on the top that enables smooth liquid flow for liquid cell studies and controllable lift off materials after deposition.

See supplementary material for Figs. S1 and S2. Figure S1 shows a schematic illustration of the chip for *in situ* TEM heating experiments. Figure S2 shows the metallic structure with a thin SiN layer after plasma etching and without SiN after vaporized HF treatment.

The authors gratefully acknowledge ERC Project NEMinTEM 267922, the STW Perspectief project UPON and European Union's Horizon 2020 research and innovation programme under Grant Agreement No. 696656.